\documentstyle[aps,prl,epsf,preprint]{revtex}
\def\la{\mathrel{\mathpalette\fun <}}
\def\ga{\mathrel{\mathpalette\fun >}}
\def\fun#1#2{\lower3.6pt\vbox{\baselineskip0pt\lineskip.9pt
        \ialign{$\mathsurround=0pt#1\hfill##\hfil$\crcr#2\crcr\sim\crcr}}}

\begin{document}


\title{\vskip-2.5truecm{\hfill \baselineskip 14pt {
\small  FERMILAB-Pub-96/170-A}\vskip .1truecm}
\vskip 0.1truecm {\bf Preheating and vacuum metastability in Supersymmetry
}}


\author{Antonio Riotto\thanks{
 riotto@fnas01.fnal.gov}$^{(1)}$, Esteban Roulet\thanks{
roulet@susy.he.sissa.it}$^{(2)}$ and Iiro  
Vilja\thanks{vilja@newton.utu.fi}$^{(3)}$}

\address{$^{(1)}${\it NASA/Fermilab Astrophysics Center, Fermi
National Accelerator Laboratory,\\Batavia, IL 60510, USA}}

\address{$^{(2)}${\it International School for Advanced Studies, SISSA-ISAS,\\
Via Beirut 2/4, I-34014, Miramare, Trieste, Italy}}

\address{$^{(3)}${\it Department of Physics, University of Turku,\\
Finland 20014, University of Turku}}

\maketitle


\begin{abstract}
\baselineskip 12pt
The constraints imposed by the requirement that the scalar potential
of supersymmetric theories does not have unbounded directions and
charge or color breaking minima deeper than the usual electroweak
breaking minimum (EWM) are significantly relaxed if one just allows
for a metastable EWM but with a sufficiently long lifetime. For this
to be acceptable one needs however to explain how the vacuum state
reaches this metastable configuration in the first place. We discuss
the implications for this issue of the inflaton induced scalar masses,
of the supersymmetry breaking effects generated during the preheating
stage as well as of the thermal corrections to the scalar potential
which appear after reheating.  We show that their combined effects
may efficiently drive the scalar fields to the origin, allowing
them to then evolve naturally towards the EWM.
\end{abstract}

\thispagestyle{empty}

\newpage
\pagestyle{plain}
\setcounter{page}{1}
\def\beq{\begin{equation}}
\def\eeq{\end{equation}}
\def\beqa{\begin{eqnarray}}
\def\eeqa{\end{eqnarray}}
\def\tr{{\rm tr}}
\def\x{{\bf x}}
\def\p{{\bf p}}
\def\k{{\bf k}}
\def\z{{\bf z}}
\baselineskip 20pt
One of the peculiar aspects of supersymmetric theories is that they
have a very extended scalar sector. In addition to the two
complex Higgs doublets of the minimal supersymmetric standard 
model, the  scalar
partners of all leptons and quarks are also present, making the
structure of the scalar potential of the theory very rich.

The fact that these scalars may carry lepton or baryon number, be
electrically charged or even colored makes in principle possible the
existence of minima of the scalar potential where the symmetries
associated to these charges are spontaneously broken. These so-called
charge and color breaking (CCB) minima are of course to be avoided,
and the requirement of having the usual electroweak breaking minimum (EWM) 
deeper than the CCB ones puts important constraints on the parameters,
like the soft scalar masses $\widetilde{m}$, the gaugino masses $M$,  the  
bilinear $B$ and the  trilinear $A$ soft
breaking terms. Particularly
dangerous are the many directions in field space giving vanishing
$D$--terms, {\it i.e.} the so called $D$--flat directions, since for them the
renormalisable 
potential can become unbounded from below for large field values  due
mainly to the effects of the trilinear couplings or to the effects of
radiative corrections to the potential. Depending upon the particular  
directions, the CCB condensate $\varphi$ may assume values of the same
order of  the weak scale or (much) larger. 

Detailed analysis of the constraints imposed by the requirement of
having a well behaved scalar potential have been performed by many authors,  
resulting
in restrictive bounds on the parameter space of supersymmetric
theories \cite{Frere}--\cite{stru96}. A complete
analysis of all the potentially dangerous directions in the field
space of the MSSM has been recently  carried
out in ref.\cite{CCB}, where it was shown that   extensive regions of the   
parameter space $(\widetilde{m},M,B,A)$ become forbidden. The
constraints  to  
avoid CCB true minima are sometimes so strong that for instance
 the whole parameter
 space is excluded in the  
particular case in which the dilaton is the source of supersymmetry breaking
\cite{casas}.

It has however been pointed out that imposing that the
EWM  be the deepest one may actually be exceedingly restrictive,
since a deeper minimum or an unbounded direction is harmless
as long as the lifetime of the now metastable EWM is longer
than the age of the Universe \cite{Roulet,color}.
In order to study the metastability of the EWM, 
one has to consider the quantum tunneling at zero temperature 
from one vacuum to the deeper one and also  take into
account the possibility of producing the transition by thermal effects
in the hot early Universe. As a result, it turns out that in many
cases  the charge and color breaking 
effects are in practice not dangerous, and hence 
these kind of considerations can 
significantly relax the bounds obtained under the
requirement of absolute stability of the EWM of the
potential \cite{Roulet}.

However, in order for these relaxed bounds to be reliable,
one has to explain how does the Universe manage to reach the color conserving  
minimum $\varphi=0$
in the first place. Indeed, 
it is  possible    that the color breaking condensate  is  left initially far  
from the 
origin, {\it e.g.} at an early epoch near the end of inflation, and
may then roll towards some  unbounded direction or CCB minimum  
before reaching  the EWM.

The discussion of the conditions under
which the Universe may be assumed to populate the EWM at early stages will be  
the main issue of this
paper. In this respect, the determination of the initial conditions
may depend crucially on the details of the inflationary epoch and of the
subsequent period of inflaton decay and thermalisation. 

As we will show,  the new idea of the preheating stage \cite{explosive}  
produced by
the resonant inflaton decay can be helpful to push an initially 
nonvanishing   color breaking condensate to the 
 origin $\varphi=0$   and hence, when the temperature of the Universe drops  
down and the electroweak phase transition occurs,   to the EWM  
where it may
then remain trapped even if this state is actually metastable, 
thus relaxing the bounds
obtained imposing that the EWM be the deepest one in the parameter space.

There are essentially three main contributions to the scalar potential which  
can
help the vacuum state to naturally evolve towards the
origin and avoid becoming trapped in a deeper but far away CCB minimum:

\noindent $i$) Scalar mass terms, $\Delta {m}^2_H$, 
proportional to the Hubble parameter
induced by the inflaton potential energy \cite{moduli,dine}.

\noindent $ii$) Temperature dependent masses,  $\Delta
{m}^2_T$,  arising after the
thermalization of the Universe.

\noindent $iii$) Supersymmetry breaking effects produced by the large
scalar condensates generated during the preheating stage  
\cite{linderiotto,dvaliriotto} according to the new theory of reheating  
\cite{explosive}. We will indicate the corresponding corrections to the soft breaking masses by $\Delta {m}^2_{{\rm pr}}$. 

Regarding the first one, it is known that the potential energy
responsible for inflation induces a contribution to the soft masses of the form 

\beq
\Delta {m}^2_H=c \:H^2,
\eeq
where $H$ is the Hubble parameter ($H\sim 10^{13}$~GeV during inflation) 
and  $c$ is a constant of  
order unity. The presence of such a contribution is a signal of supersymmetry  
breaking during inflation. 

For the minimal form of the  K\"ahler
potential leading to canonical kinetic terms, $c$ is positive and one   may  
expect that during inflation the color breaking condensate 
$\varphi$ is driven  
exponentially to the origin even if it starts from very large values  
$\varphi\sim M_P$. In such a case, the considerations discussed in  
\cite{Roulet,color} may be applied. Nonetheless,  the constant $c$ may
vanish \cite{murayama} or even be   
negative in more general
models \cite{dine}. Negative values of $c$ 
along particular flat directions may in fact be required for the  
Affleck-Dine mechanism of baryogenesis \cite{ad} to work, so as to set the  
scalar fields to large initial values. 

We will hence focus hereafter in the more problematic 
case $c<0$, and study the 
evolution of the fields along flat directions which, at zero
temperature, present an undesirable minimum $\varphi_0$ at large field
values (this could typically be the case for an unbounded direction of
the renormalisable potential lifted by a non-renormalisable
interaction term, see below), and then comment on the `non--flat'
case. 
 
The scalar potential   
for a $D$--flat direction during inflation and in the presence of
a non-renormalisable superpotential
of the type $W_{NR}=(\lambda/n M^{n-3})\varphi^n$ can
be represented as follows
\cite{dine}:
\begin{equation}
V(\varphi)  =     c \:H^2|\varphi|^2  + \lambda^2{|\varphi|^{2n-2}\over    
M^{2n-6}},
\end{equation}
where $M$ is some large mass scale such
as the GUT or Planck
mass and  $n$ is some integer which may take values larger than 3.
  This leads to symmetry breaking
with $ \langle\varphi\rangle \sim (H M ^{n-3}/\lambda)^{1/n-2}$.

In the old theory of reheating \cite{old}, after the inflationary stage the  
inflaton field $\phi$  starts oscillating around the minimum of its potential
and the Universe soon  becomes  matter-dominated (assuming a quadratic
inflaton potential). In such a stage, $H\propto a^{-3/2}$, 
 $a(t)$ being the  expansion scale factor, and 
the minimum $\langle\varphi(t)\rangle\propto H^{1/n-2}$ decreases
accordingly. 
 At the time $t_d\sim \Gamma_\phi^{-1}$, the inflaton  
decays with a decay width $\Gamma_\phi$ and the inflaton  energy is released  
under the form of light relativistic particles which thermalize and 
reheat the  
Universe up to a temperature
$T_R\simeq 10^{-1}\sqrt{\Gamma_\phi M_P}$. At this epoch, 
if the renormalisable terms of the superpotential can still be
neglected, the flat direction minimum will be at 
\beq
\langle\varphi\rangle_R\simeq 
\varphi_d\equiv \left( {\Gamma_\phi M^{n-3}\over \lambda}
\right)^{\frac{1}{n-2}},
\eeq
which for $n=4$ and $M=M_P$ is of the order of $10\, T_R/\sqrt{\lambda}$.
Note that, for $c<0$, the effect of the inflaton induced mass is
always to shift the minimum of the potential to values larger than the
zero temperature minimum $\varphi_0$, and hence for the previous
reasoning to
apply  the value of $\varphi_d$ obtained above should be  larger than
$\varphi_0$. This may not be the case 
if the renormalisable terms are actually non negligible, in which case 
 one may just assume that $\langle\varphi\rangle_R
\simeq\varphi_0$ at the reheating time. 

After reheating, the negative correction $\Delta{m}^2_H$ disappears 
because the inflaton energy has been reduced to zero and the 
effective potential $V(\varphi)$ consists now of the zero 
temperature piece  and, for not too large  values of the field, the
thermal effects may provide a new important contribution.
Indeed, the fields directly  
interacting with $\varphi$ acquire masses $\simeq
g\langle\varphi\rangle$, where $g$ is their coupling to the flat
direction. Hence, after reheating
 they may become  excited by thermal effects and induce 
a finite temperature correction of order $g^2\:T^2\:\varphi^2$
to the effective potential 
as long as $g\langle\varphi\rangle_R\la T_R$. 
If the renormalisable terms are small, so that
$\langle\varphi\rangle_R\simeq\varphi_d>\varphi_0$, this condition
translates into 
\beq
\label{gbound}
 g \la  0.1\left({\Gamma_\phi\over
M}\right)^{\frac{n-4}{2n-4}}\sqrt{M_P\over M}\lambda^\frac{1}{n-2}. 
\eeq
As an example, this implies that  $g$ is to be smaller than 
$\sim 0.1\sqrt{\lambda M_P/M}$ for $n=4$, which is not difficult 
to satisfy. However, 
for $n\gg 4$ the vacuum expectation value (VEV)  $\langle\varphi
\rangle_R$ turns out to be quite
large ($\sim M$) and one needs then $ g\la T_R/M$, 
a requirement which could be very stringent. 
If the condition in Eq.~(\ref{gbound}) is satisfied so that  the flat direction
 soft breaking mass receives a temperature dependent contribution  
$\Delta {m}^2_T\sim g^2T^2$ due to the effect of the light particles
coupled to $\varphi$, this new contribution can help to drive the
field to the origin. Indeed, 
 if just after reheating this thermal 
correction dominates the scalar
potential for $\varphi\la\langle\varphi\rangle_R$, it  may completely
 drag the CCB   condensate to the origin. The detailed conditions
under which this happens however depend crucially on the actual value
of $g$, which cannot be too small for $\Delta m_T^2$ to be sizeable,
and on the strength of the renormalisable terms in the  potential which
could compensate the thermal effects and  push the
field away from the origin.

The same problem is present if the renormalisable terms are sizeable 
at reheating 
and make the CCB $\varphi_0$  much larger than both  $\varphi_d$ and
$T_R/g$.
 Again,  the time-dependent
minimum  
 relaxes towards $\varphi_0$ to remain trapped  
there since there are no significant temperature induced corrections.
 This is unacceptable and the parameter space of the  
theory leading to this kind of situation should be eliminated. 
In this case, only if $T_R$ is large enough (its lower bound depending upon
 the different situations) it may happen 
that the thermal effects push
the fields towards the origin and allow them to  
evolve later towards the EWM. On the other hand,
 the  CCB minima whose present VEV's are of the order of the 
weak scale (such as those  along the $D$--flat direction $\varphi=H_2^0=
\widetilde{t}_L=\widetilde{t}_R$ or along the direction $\varphi=
\widetilde{t}_R$ with negative soft breaking mass $\widetilde{m}_R^2$
\cite{color,mariano}) are very unlikely to be ever  populated in the 
early Universe. Indeed, it is  quite reasonable to expect the reheating 
temperature after inflation to be much larger than the weak scale,
making  thermal corrections  very efficient
in driving the color breaking condensate to the 
origin. Hence, in this case
 the regions of parameter space for which the lifetime
of the EWM is sufficiently large can be considered as cosmologically 
acceptable and do not  pose any problems to the consistency of the 
theory. 

In the rest of the paper we will therefore concern ourselves with the
case of CCB minima whose present VEV's are (much) larger than the weak
scale and    investigate how a preheating stage after inflation
may modify the picture arising from the old theory of reheating that
we sketched above. 
Indeed, it has been recently pointed  
out that the inflaton may decay  explosively just at the end of inflation
through the phenomenon of parametric resonance \cite{explosive},
leading to a situation quite 
different from the one predicted in the old theory of reheating which
could also be helpful in driving the scalar fields towards the
origin.

According to this new scenario, a significant fraction of the inflaton
energy 
is released in the form of bosonic inflaton decay
products, whose occupation number is extremely large, and may have
energies much smaller than the temperature that would have been
obtained by an instantaneous conversion of the inflaton energy density
into radiation.
Since it requires several scattering times for the low-energy decay
products to form a thermal distribution, it is rather reasonable to
consider the period in which most of the energy density of the
Universe was in the form of the nonthermal   quanta produced by
inflaton decay as a separate cosmological era, dubbed as preheating to
distinguish it from the subsequent  stages  of particle decay and
thermalization which can be described by the techniques developed in  
\cite{old}.
Several aspects of the theory of explosive reheating have been studied in the
case of slow-roll inflation \cite{noneq} and first-order inflation
\cite{kolb}.
One of the most peculiar   aspects of the stage of preheating is the
possibility of  nonthermal phase transitions with symmetry restoration
\cite{KLSSR,tkachev,rt} driven by extremely large quantum corrections induced  
by particles generated during the stage of preheating. 

The key observation is that fluctuations of scalar fields produced 
at preheating may be  so large 
that they  
can break supersymmetry much strongly than
inflation itself \cite{linderiotto,dvaliriotto}. This may happen since
parametric resonance is a phenomenon characteristic of bosonic particles and
the resonant decay into fermions is inefficient because of Pauli's 
exclusion principle. Therefore, during the preheating stage the
Universe is populated only by a huge number of bosons and the 
occupation numbers of bosons and fermions of the same supermultiplet 
coupled to the inflaton are unbalanced. Supersymmetric cancellations 
between diagrams involving bosons and fermions are no longer operative
 at this stage and large loop corrections appear. The preheating stage
 is then intimately associated to strong supersymmetric breaking and 
large fluctuations may lead  to symmetry restoration along
flat directions of the effective potential even in the theories where 
the usual
high temperature corrections   are exponentially suppressed. Hence, the  
curvature along $D$--flat directions during the preheating may be much larger  
than the inflaton induced effective mass.
This may render 
the details of the effective potential along $D$--flat directions 
during inflation almost  
irrelevant as far as the initial conditions of the condensates along 
these  directions is concerned, and may 
have a profound impact on the fate of CCB minima. 

Let us take as an illustrative case the one usually considered in which
the inflaton decays into a field $\chi$, and assume that $\chi$ is
coupled to the flat direction $\varphi$, with the potential
\beq
V= \frac{M_\phi^2}{2}\phi^2+g_\phi^2\phi^2|\chi|^2+ g_\varphi^2 |\varphi|^2|\chi|^2+ g_\chi^2
|\chi|^4 ,
\eeq
where  $M_\phi\sim 10^{13}$ GeV in order for the density
perturbations generated during
the inflationary era to be consistent with COBE data \cite{cobe}
and where the last term could be for instance a $D$--term self--coupling of
$\chi$ if this field is not a gauge singlet.

Inflation occurs during the slow rolling of the inflaton field 
until it reaches a  value $\phi_0\sim M_P$. Then it starts 
oscillating with an initial amplitude $\phi_0$ and a significant fraction of the initial 
energy density $\rho_\phi\sim M_\phi^2\phi_0^2$ is transferred 
to bosonic $\chi$-quanta in the regime of parametric resonance.
Let us postpone the discussion of the conditions under which 
parametric resonance occurs and go directly to the   main 
observation of this paper. 

At the end of the broad resonance regime the inflaton field 
drops down to $\phi_e\sim 10^{-2} M_P$ and the Universe is 
filled up with $\chi$-bosons
with a typical energy $E_\chi\sim 0.2\sqrt{g_\phi M_\phi M_P}$ 
and a large occupation number $n_\chi/E_\chi^3\sim 1/g_\phi^2$. 
The amplitude of the field fluctuations produced at this stage 
is very large \cite{explosive,KLSSR}
\begin{equation}
\label{r1}
\langle\chi^2\rangle\sim 5\times 10^{-2} g_\phi^{-1} M_\phi M_P.
\end{equation}

It is exactly the incredibly large value\footnote{
If the energy of the inflaton field after preheating were 
instantaneously thermalized, a much smaller value of $\langle
\chi^2\rangle$ would have been obtained, $\langle\chi^2\rangle
\sim 10^{-4} M_\phi M_P$. }
 of $\langle\chi^2\rangle$ 
that leads to nonthermal symmetry restoration during the preheating 
stage, drives strong supersymmetry breaking and is responsible for 
the  additional contribution to the effective mass along the 
$D$--flat direction
\begin{equation}
\Delta {m}^2_{{\rm pr}}\sim g_\varphi^2\langle\chi^2\rangle\sim 10^{-1}\:\frac{g_\varphi^2}{g_\phi}\:M_\phi M_P.
\label{mpr}
\end{equation}

The curvature of the effective potential along the $D$--flat direction
becomes large and positive and the symmetry is restored if
$\Delta{m}^2_{{\rm pr}}> \delta |c| H^2$, where $\delta\sim 
10^{-1}$ parametrizes the fraction of the energy density still 
stored in the inflaton field after the end of the preheating stage.

Since at the end of preheating $H\sim 10^{-2} M_\phi$, this happens if
\begin{equation}
\label{con}
g_\varphi^2\ga 10^{-9}\: g_\phi.
\end{equation} 

Let us note that the scalar background corrections to the  masses in
Eq.~(\ref{mpr}) are a consequence of  forward scattering
processes which do not alter the distribution function of the particles
traversing the $\chi$ background, but simply modify their dispersion
relation.  Since the
forward scattering rate is usually larger than the large-angle
scattering rate responsible for establishing the thermal distribution,
this contribution can be present 
even before the initial nonequilibrium distribution function of the 
$\chi$ particles relaxes to its thermal value. Actually, in realistic
models thermalization  typically takes 
a lot of time and the value of $\langle\chi^2
\rangle$ after the Universe reheats is much smaller than (\ref{r1}). 
Notice also that Eq.~(\ref{con}) is approximately the same condition which enables 
the interaction rate of the forward scatterings to be large  enough, 
i.e. $\Gamma_{{\rm fs}}\simeq \sqrt{\Delta{m}^2_{{\rm pr}}}>H$,  for producing
the correction to the soft breaking masses and  lifting the curvature of the 
$D$--flat direction. 

Hence, we have shown that  if 
the Universe spent an intermediate stage of preheating, it is
possible 
that the initially nonvanishing color breaking condensate was 
dragged to the origin
as a consequence of  strong supersymmetry breaking and nonthermal 
symmetry restoration. 
If so, 
even though a deeper CCB minimum may be present in our (almost) 
supersymmetric world, the considerations of metastability of 
the EWM mentioned at the begining may safely be applied in order to
explore the allowed parameter space of the theory.
All these considerations apply provided   the condition (\ref{con}) 
is fulfilled so that even a negative inflaton induced mass  can be
compensated by the preheating effects. 
To get the feeling of the restrictions this  condition poses, let  
consider  the $D$--flat direction 
$\widetilde{u}_R^1=\widetilde{s}_R^2=\widetilde{b}_R^3\equiv \varphi$
discussed  in refs.~\cite{Falk,Roulet}. 
If $\chi$ is not one of the fields labelling the flat direction,  
the coupling $g_\varphi$ should  be identified with a Yukawa
coupling. 
If the inflaton couples to  $\widetilde{b}_L$ or $H_1^0$ then 
$g_\varphi=h_b$ and the inequality  (\ref{con}) is easily 
fulfilled. A more concrete example may be provided in the 
case in which chaotic inflation is driven by a right-handed 
sneutrino $\widetilde{\nu}^c$ \cite{right}. The latter may 
decay by parametric resonance into sleptons and Higgses through 
the coupling in the superpotential $\delta W=
h_\nu \nu^c L H_2$. The coupling $g_\varphi$ is then identified 
with the Yukawa coupling of the $u$-quark and condition (\ref{con}) 
is satisfied if $g_\phi\equiv h_\nu\la 10^{-1}$. 
These simple estimates indicate that, even though  
the validity of the results depends upon the details of the theory, 
the  mechanism of nonthermal symmetry  
restoration along CCB $D$--flat direction may offer a concrete
explanation of why does the Universe manage to sit on the color
conserving   minimum $\varphi=0$ at early stages. 

Once the condensate is driven to the origin, it is likely to stay 
there even at later epochs. Indeed, 
after the end of the broad resonance the Universe is radiation 
dominated and the value of $\langle\chi^2\rangle$ decreases 
approximately as $a^{-2}\sim t^{-1}$. As a result,
$\Delta{m}_{{\rm pr}}^2$  decreases as $t^{-1}$, while $H^2$ 
scales like $t^{-2}$.  Hence    
the condition
$\Delta{m}_{{\rm pr}}^2>\Delta m^2_H$  continues to be satisfied 
after the end of the broad resonance stage 
 and the condensate remains
 trapped at the origin. 
This remains true even if there is some intermediate stage when the 
Universe suffers a  matter dominated period, e.g. if the   residual 
energy density stored in the form of  inflaton oscillations, which at 
the end of the broad resonant regime was reduced to a fraction
$\delta$ of the total energy density
but however decreases more slowly than the radiation,
starts dominating again the 
energy density of the Universe.

When finally thermalization of the $\chi$ background 
occurs at the time $\tau$,   the Universe 
is reheated  to a temperature $T(\tau)\simeq T_R\sqrt{\tau_0/\tau}$ 
\cite{linderiotto}, where $T_R\simeq 
10^{-2}\sqrt{M_\phi M_P}$ is the temperature which the system would 
have reached if thermalization occured instantaneously after the 
preheating stage and $\tau_0\sim 10^8/M_P$ is the time at the end of 
the broad resonance regime\footnote{It is worthwhile mentioning that 
particles popping out  from the thermalization processes are not 
guaranteed to be in thermal equilbrium if 
the reheating temperature is somewhat higher than $10^{14}$ GeV \cite{out}.}.
At this moment the $D$--flat direction is lifted by the term $\sim 
T^2(\tau)\varphi^2$ arising from the thermal effects associated to the
light particles coupled to $\varphi$ (remember that now
$\langle\varphi\rangle\simeq 0$). Since  $T^2(\tau)\gg H^2(\tau)$, 
we may conclude that the CCB condensate will remain trapped near the
origin by thermal effects and  relax  towards the EWM at later stages,
 unless the coupling $g_\varphi$ is too small to fulfill Eq.~ (\ref{con}).
 
We now  further check the applicability of our predictions and discuss
in more detail the requirements for having parametric resonance. 
For the resonance to be `broad',  the
coupling of the inflaton to $\chi$ should not be too small, $g_\phi
\phi_e>M_\phi$,  where $\phi_e\sim 10^{-2} M_P$ at the end of 
preheating. This leads to $g_\phi>10^{-4}$. Notice that the flatness 
of the inflaton  potential during  
inflation  is preserved for such large values of couplings $g_\phi$ by  
supersymmetric cancellations \cite{dvaliriotto}.
The contribution $g_\varphi\langle\varphi\rangle\sim g_\varphi(H
M^{n-3}/\lambda )^{1/n-2}$ to the effective mass of the field $\chi$
 induced by the condensate $\langle\varphi\rangle$ at the end of the 
preheating stage should be smaller than the typical energy of the 
decay products $E_\chi$. For $n=4$ and $M=M_P$, this translates 
into the bound $g_\varphi^2\la g_\phi$. 

Interactions which reduce the number of $\chi$ quanta produced during
the inflaton decay 
may remove the decay products of the inflaton and take the system 
 away from the resonance
shell, thus stopping the parametric resonance stage 
(self-interactions of the $\chi$-field instead just keep particles 
inside the resonance shell). This can be avoided in two different 
ways. Either the scatterings are suppressed by kinematical reasons, 
namely if the nonthermal plasma mass of the final states is larger 
than the typical energy of the $\chi$'s. Otherwise,  and more 
realistically, scatterings occur, but are slow enough that they 
do not terminate the resonance. This may happen if the interaction 
rate $\Gamma\sim n_\chi\sigma$, where $\sigma\propto
\alpha^2/E_\chi^2$ 
denotes the scattering cross section, is smaller than the typical 
frequency of oscillations at the end of the preheating stage $\sim 
g_\phi\langle \chi^2\rangle^{1/2}$. This translates into the bound 
$\alpha\la 10 \:g_\phi$. 

If $\chi$ particles self-interact strongly, e.g. if the $\chi$-field 
is not a gauge singlet and $g_\chi$ is a $D$--term self--coupling
larger than $g_\phi$, some particular care is needed \cite{rt,inprep}. 
The resonance stops when the value of the plasma mass of the
$\chi$-field 
induced by nonthermal effects, $m_\chi^2\sim g_\chi^2 \langle\chi^2
\rangle$, 
becomes of the order of $\sqrt{g_\chi n_\chi E_\chi}$, i.e. when 
 the fluctuation $\langle\chi^2\rangle$ becomes  $\sim 5\times 
10^{-2} g_\chi^{-1} M_\phi M_P$ and $n_\chi/E_\chi^3\sim
g_\chi^{-2}$. 
In such a case, all the conclusions drawn so far remain 
qualitatively inaltered, even though our picture changes from the 
quantitative point of view. For instance, nonthermal symmetry 
restoration along the $D$--flat direction now occurs if $g_\varphi^2
> 10^{-8} \delta\, g_\chi$ and the resonance is not stopped by production of 
different quanta if $g_\phi>10^{-2} g_\chi^3$, where we have assumed 
$\alpha\sim g_\chi^2$. Thermalization is expected to occur earlier  
than in the weak coupling limit and the thermalization temperature 
$T(\tau)$ is consequently higher. 

The discussion above makes clear that the validity of the scenario
discussed is based on several assumptions about 
the structure
of the theory and on relations between various coupling constants. 
These assumptions should be explicitly checked when dealing with a 
completely realistic model of inflation. It is anyhow encouraging
that the recent progress in the theory of reheating may help in 
removing one of the dangers present in any 
supersymmetric theory, namely the presence of unwanted CCB minima 
which make the EWM metastable. We have shown that the combined 
effects of the  strong supersymmetry breaking during the preheating 
stage and the thermal corrections appearing after the reheating 
may provide a concrete explanation of why the origin of the field
space was populated at early times, allowing the condensate to then 
evolve naturally towards the metastable EWM at temperatures around 
the weak scale. Since even in the presence of a deeper CCB minimum 
the lifetime of the EWM is often longer than the age of the
Universe,  in these cases the 
charge and color breaking effects may actually pose no problems.

\acknowledgments
We would like to thank Andrei Linde and  Leszek Roszkowski for 
discussions. AR was 
supported in part by the DOE and by NASA (NAG5-2788) at Fermilab. IV thanks the  
Theoretical Astrophysics Group at Fermilab for the kind hospitality. 

\def\MPL #1 #2 #3 {{\em Mod.~Phys.~Lett.}~{\bf#1}\ (#2) #3 }
\def\NPB #1 #2 #3 {{\em Nucl.~Phys.}~{\bf B#1}\ (#2) #3 }
\def\PLB #1 #2 #3 {{\em Phys.~Lett.}~{\bf B#1}\ (#2) #3 }
\def\PR  #1 #2 #3 {{\em Phys.~Rep.}~{\bf#1}\ (#2) #3 }
\def\PRD #1 #2 #3 {{\em Phys.~Rev.}~{\bf D#1}\ (#2) #3 }
\def\PRL #1 #2 #3 {{\em Phys.~Rev.~Lett.}~{\bf#1}\ (#2) #3 }
\def\PTP #1 #2 #3 {{\em Prog.~Theor.~Phys.}~{\bf#1}\ (#2) #3 }
\def\RMP #1 #2 #3 {{\em Rev.~Mod.~Phys.}~{\bf#1}\ (#2) #3 }
\def\ZPC #1 #2 #3 {{\em Z.~Phys.}~{\bf C#1}\ (#2) #3 }

\end{document}